\documentclass[12pt]{article}

\usepackage{scicite}
\usepackage{graphicx}
\usepackage{xcolor}
\usepackage{times}
\usepackage{upgreek,xspace}
\usepackage{amssymb}
\usepackage{bm}
\usepackage{ulem}
\usepackage{comment} 
\usepackage{bibunits}
\newenvironment{sciabstract}{%
\begin{quote} \bf}
{\end{quote}}

\newcounter{lastnote}

\title{Self-organized Lasers of Reconfigurable  Colloidal Assemblies}

\author
{Manish Trivedi,$^{1\dagger}$ Dhruv Saxena,$^{2\dagger}$ Wai Kit Ng,$^{2\dagger}$  Riccardo Sapienza,$^{2\ast}$ Giorgio Volpe$^{1\ast}$\\
\\
\normalsize{$^{1}$Department of Chemistry, University College London, 20 Gordon Street,}\\
\normalsize{London WC1H 0AJ, United Kingdom}\\
\normalsize{$^{2}$The Blackett Laboratory, Department of Physics,}\\
\normalsize{Imperial College London, London SW7~2BW, United Kingdom}\\
\\
\normalsize{$^\dagger$These authors contributed equally}\\
\normalsize{$^\ast$To whom correspondence should be addressed; E-mail: r.sapienza@imperial.ac.uk; g.volpe@ucl.ac.uk}
}

\date{}

\begin{document} 

\maketitle 

\begin{sciabstract}
Biological cells self-organize into living materials that uniquely blend structure with functionality and responsiveness to the environment. The integration of similar life-like features in man-made materials remains challenging, yet desirable to manufacture active, adaptive and autonomous systems. Here we show the self-organization of programmable random lasers from the reversible out-of-equilibrium self-assembly of colloids. Random lasing originates from the optical amplification of light undergoing multiple scattering within the dissipative colloidal assemblies and therefore is crucially dependent on their self-organization behavior. Under external light stimuli, these dynamic random lasers are responsive and present a continuously tunable laser threshold. They can thus reconfigure and cooperate by emulating the ever-evolving spatiotemporal relationship between structure and functionality typical of living matter. 
\end{sciabstract}


\newpage


Self-organization is the spontaneous emergence of structure and coordination from elementary units on larger scales than those defining the individual components \cite{Whitesides2002}. Over billions of years of evolution, living matter has mastered this process to form dynamic functional structures at various scales, from biomolecules, to cell membranes, to tissue. At its heart lies the non-equilibrium nature of biological cells that can aptly and autonomously convert available energy to function and perform actions.

The desire to incorporate life-like dynamics into self-assembly methods of artificial materials has driven a vast scientific effort in the hope to manufacture materials that are active, adaptive and autonomous \cite{AnimateMaterials}. In colloidal science, active colloids (colloids capable of performing actions after energy conversion) have come to the fore due to their ability to engage in dissipative self-assembly \cite{Bechinger2016, mallory2018active}, thus enabling to emulate self-organization in living matter \cite{palacci2013living, Pince2016, khadka2018active, Bechinger2019} and to implement microscopic active metamachines and mechanical devices \cite{maggi2016self, yuan2018self,aubret2018targeted}. Despite their potential, these dissipative colloidal assemblies have yet to demonstrate the intricate dynamic relationship between structure and functionality shown by most living assemblies, which can reshape their structure to perform different functions in response to external cues.

Because of the ease of synthesis with sizes comparable to the wavelengths of visible light, (non-active) colloids have often been the building blocks of choice for photonic materials and devices with optical properties defined by their fixed topology and spatial correlations \cite{correlatedreview,bioinspiredcolloids}. Adding optical gain to these static photonic assemblies can trigger lasing \cite{DeterminingRL}. In disordered assemblies, random lasing \cite{Wiersma2008,DeterminingRL} has been observed in solid photonic glasses \cite{PhotonicGlassRL}, titania (${\rm TiO_2}$) colloidal systems \cite{lawandy1994}, semiconductor powders \cite{noginov2006solid}, and more complex geometries \cite{Feng2015}. These lasing functionalities emerging in the final assembled photonic materials are reaching technological applications (from low-coherence imaging \cite{Cao2019review} to super-resolution spectroscopy \cite{Boschetti2020}, from sensing \cite{silk2016} to even interfacing with living tissues \cite{optofluidicReview2014}), thanks to the flexibility and shape insensitivity of random lasing processes.

Yet, the formation of photonic colloidal assemblies is still overwhelmingly dominated by slow equilibrium dynamics, where the final configuration is locked and the optical functionalities static and fixed. Here we realize programmable random lasers, which self-organize from the dissipative self-assembly of colloids and show features akin to living systems, such as responsiveness, reconfigurability and cooperation.

\begin{figure}[htb]
    \centering
    \includegraphics[width=12 cm]{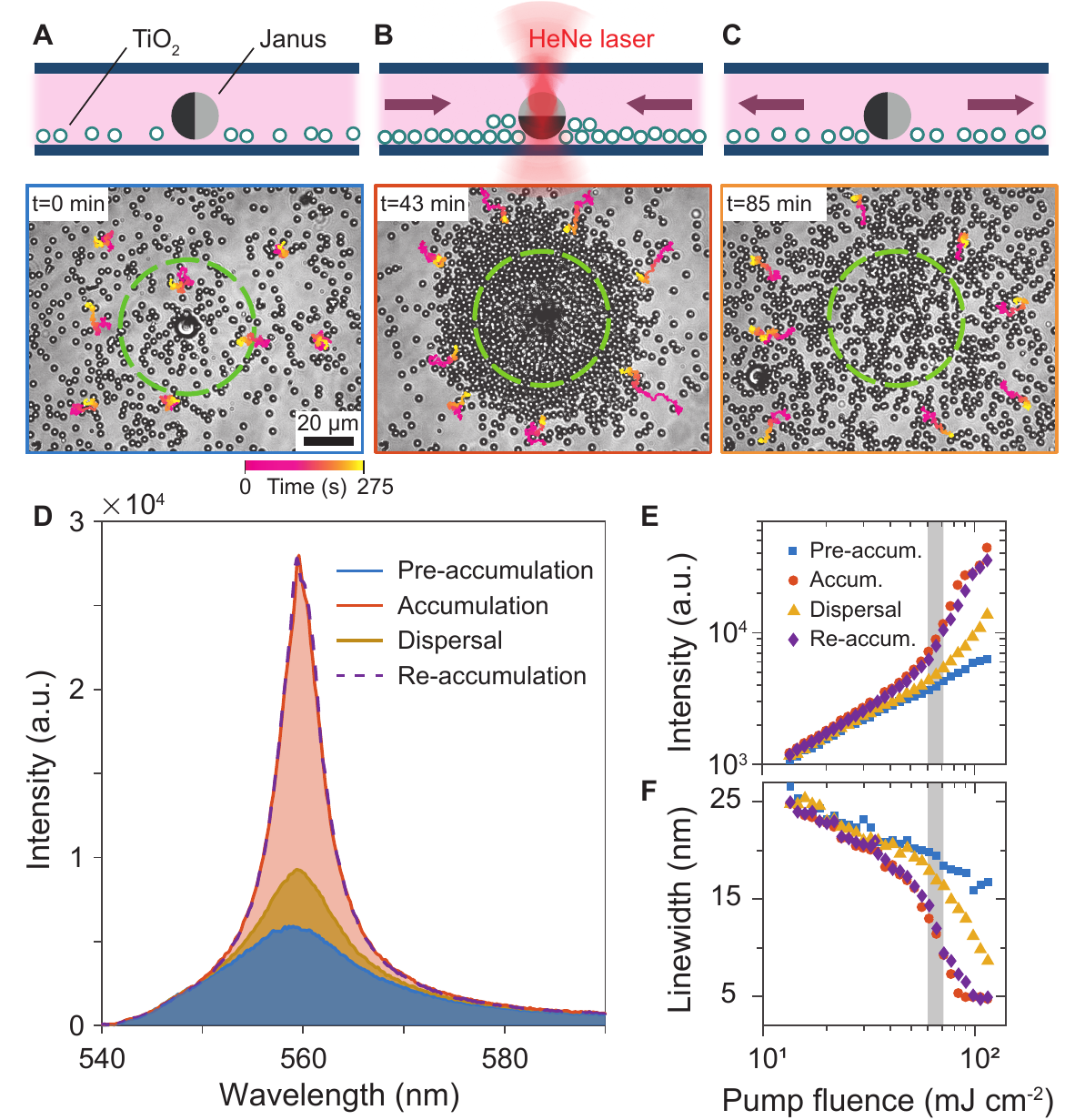}
    \caption{{\bf Reversible self-organization of colloids in a programmable random laser}. ({\bf A}) A light-absorbing Janus particle in a TiO$_2$ and laser dye (rhodamine 6G) colloidal dispersion attracts the diffusing colloids (accumulation), which ({\bf B}) assemble in a dense cluster, when illuminated by a  HeNe laser (CW, 632.8 nm) \cite{Supplementary}. ({\bf C}) If the  HeNe laser is off, the colloids disperse. ({\bf A-C}) A few 275 second-long trajectories highlight the colloids' motion. ({\bf D}) Lasing is observed upon optical pumping (400 ps laser pulses, 532 nm, pump fluence 100 mJ cm$^{-2}$, dashed area in {\bf A-C}) during accumulation, but not before (pre-accumulation) or after (dispersal). Re-accumulation after dispersal (Fig. \ref{fig:SI_Reaccumulation}) confirms lasing recovery. ({\bf E-F}) The spectra show a reversible narrowing of the emission with cluster formation, corresponding to random lasing action, confirmed by ({\bf E}) the nonlinear increase of the peak intensity as a function of pump fluence and ({\bf F}) the reduction of emission linewidth to 5 nm (below threshold at 13.5 nm). The gray-shaded regions show the pump fluence threshold for lasing ($\sim$70 mJ cm$^{-2}$).
   }
    \label{fig:Fig1}
\end{figure}

Colloids in solutions of laser dyes can scatter and amplify light that is trapped within them. When optically pumped by a high-energy laser of constant spot size and intensity~\cite{Supplementary}, lasing can emerge when the local colloidal density increases above a threshold such that light travels an average optical path long enough for net amplification to occur before leaving the medium. This is the onset of random lasing~\cite{DeterminingRL} reached by increased scattering. In Fig.~\ref{fig:Fig1}A-B, we drive the self-assembly of freely diffusing polyethylenimine-functionalized monodisperse $\rm TiO_2$ colloids of radius $R_{\rm TiO_2} = 0.915 \pm 0.03 \, {\rm \mu m}$ (Figs.~\ref{fig:SI_TiO2Diffusion}-\ref{fig:SI_Functionalization}) in an ethanol solution of a rhodamine-based dye (rhodamine 6G or B) by generating a local temperature gradient around a carbon-coated Janus particle ($R_{\mathrm S} = 4.22 \pm 0.14 \, {\rm \mu m}$) \cite{Supplementary}. Under illumination by a continuous-wave $632.8$~nm HeNe laser (Fig.~\ref{fig:SI_Setup}) \cite{Supplementary}, the Janus particle heats up because of light absorption, a $\rm TiO_2$ colloidal cluster assembles around it (Fig.~\ref{fig:Fig1}B), and the particle assumes a cap-down orientation due to its equilibrium rotational dynamics (Fig. \ref{fig:SI_JanusOrientation}). On removal of the external energy source, the colloids disperse reversibly (Fig.~\ref{fig:Fig1}C). The process of dissipative gathering can then be reiterated, leading to re-accumulation of colloids around the Janus particle (Fig.~\ref{fig:SI_Reaccumulation}). As soon as a dense cluster is formed (Fig.~\ref{fig:Fig1}B), lasing action can emerge dynamically via this phenomenon of dissipative self-organization. The lasing process is quantifiable by instantaneously measuring the emission spectra at different stages of this process (Fig.~\ref{fig:Fig1}D) by pumping optically with a $532$~nm pulsed laser (400~ps duration) at increasing pump fluence (Fig.~\ref{fig:SI_Setup}) \cite{Supplementary}. The lasing threshold is reached when the linewidth of the emission spectrum narrows to 13.5 nm, i.e. half of its initial value (Fig.~\ref{fig:Fig1}E-F). These spectra highlight the programmable optical functionality of these colloidal assemblies: lasing can be switched on/off dynamically by controlling the density and size of the cluster within the pump region (52 $\mu$m in diameter). The initial concentration  of $2 \times 10^{15}$ particles m$^{-3}$ (Fig. 1A) is too low to obtain lasing, and we only observe the broad emission characteristic of the dye fluorescence. A significantly larger, higher-density  ($12 \times 10^{15}$ particles m$^{-3}$) cluster as after accumulation and reaccumulation (Figs.~\ref{fig:Fig1}B and \ref{fig:SI_Reaccumulation}), however, shows a single narrow peak at $560$ nm with linewidth  of $\sim$5 nm. During dispersal (Fig.~\ref{fig:Fig1}C), the spectrum broadens again, thus evolving towards the initial fluorescence background. The power dependence evolution of the spectra in Fig.~\ref{fig:Fig1}E-F indicates that the cluster is lasing at a threshold power of 70 mJ cm$^{-2}$, with two clear signatures of lasing \cite{DeterminingRL}: a marked superlinear increase in the emission intensity and a significant reduction in spectral linewidth. 

\begin{figure}[htb]
    \centering
    \includegraphics[width=9.5 cm]{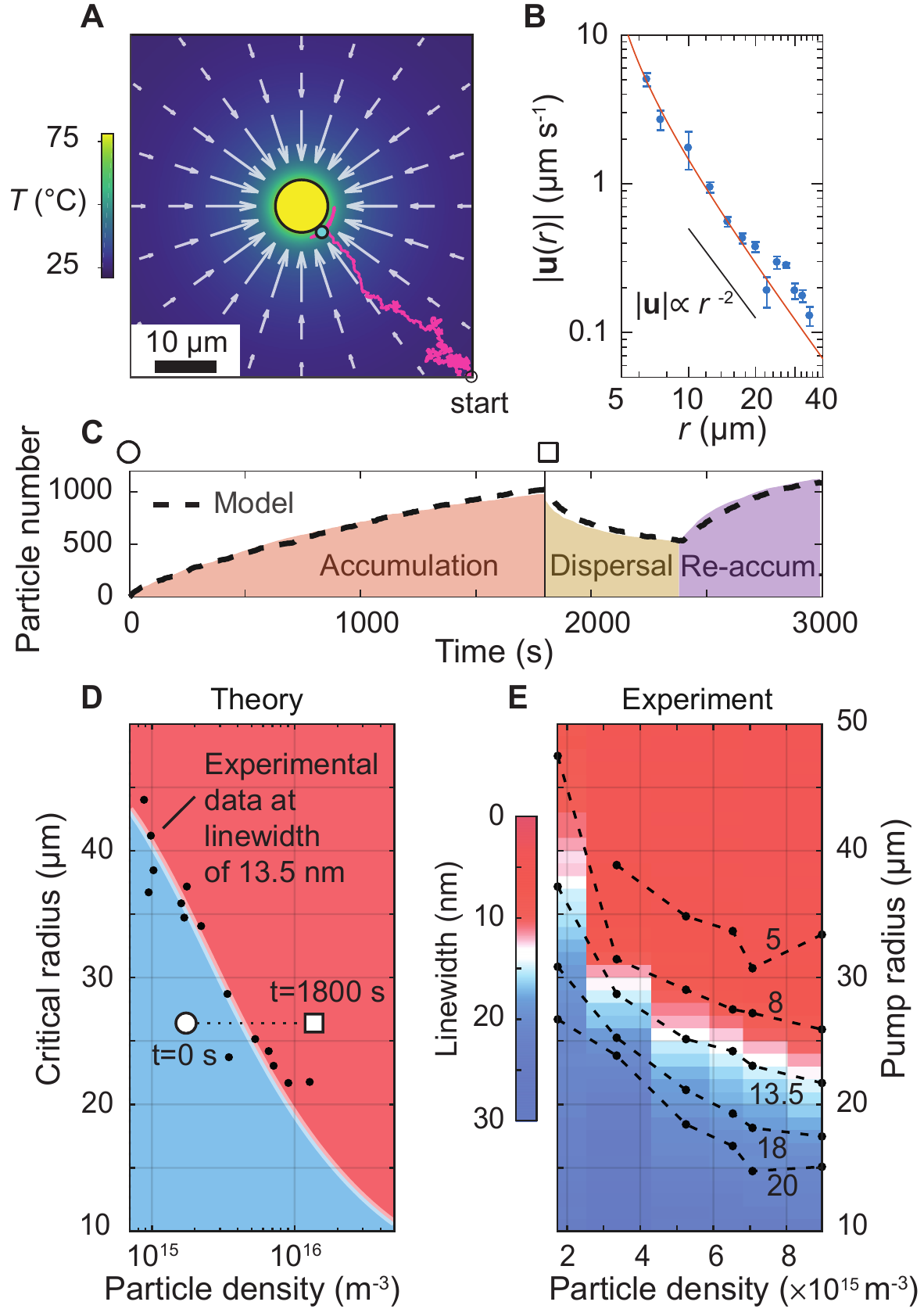}
    \caption{{\bf Dynamics of dissipative colloidal accumulation and  random lasing.} ({\bf A}) Calculated temperature profile around a heat source (yellow circle) with the corresponding thermophoretic velocity field (arrows) for a ${\rm TiO_2}$ colloid (cyan circle). 
    The over-imposed simulated trajectory (magenta line) shows how motion becomes more directed when approaching the heat source, consistent with ({\bf B}) the experimental (dots) and modeled (line) colloidal radial velocity with distance $r$ from the heat source. ({\bf C}) Time dynamics of cluster formation (shaded area: experiments; dashed line: simulation) when the heat source is on (accumulation), off (dissipation) and on again (re-accumulation). Time points (circle and square) correspond to start and end of accumulation. ({\bf D-E}) Cluster linewidth versus particle density and pump radius for ({\bf D}) random laser theory and ({\bf E}) experiments with rhodamine 6G (rhodamine B in Fig. \ref{fig:SI_RhB_Rcr}). The dots highlight experimental values  obtained ({\bf D}) at threshold and ({\bf E}) at different linewidths (values in nm in the plot) for different pump radii under a fixed pump fluence (140 mJ cm$^{-2}$). 
    The circle and square in {\bf D} show the initial and final particle densities around the heat source before and after accumulation (as in {\bf C}).
    The increase in particle density through accumulation explains the transition from below threshold (blue region) to lasing (red region). 
    }
    \label{fig:Fig2}
\end{figure}

We can interpret and quantitatively reproduce the dissipative accumulation of colloids, which leads to crossing the lasing threshold, with a two-dimensional model based on negative thermophoresis, where colloids in solution are drawn towards higher temperatures \cite{wurger2007thermophoresis}. Fig.~\ref{fig:Fig2}A shows the calculated steady-state temperature profile around the heat source (the illuminated Janus particle), which approximately decays with the inverse of the radial distance $r$ from the source \cite{Supplementary}. For the fixed HeNe laser intensity in our experiments (0.14 mW ${\rm \mu m^{-2}}$), we measured a temperature increase $\Delta T = 57 \pm 1.6 \, ^{\circ} {\rm C} $ over room temperature, corresponding to a source temperature $T_{\mathrm s} = 78 \pm 1.6 \, ^{\circ} {\rm C} $ (Fig. \ref{fig:SI_TMeasurement}). This sharp radial thermal gradient induces Marangoni forces, which drag the ${\rm TiO_2}$ particles in the radial direction determined by their Soret coefficient  \cite{wurger2007thermophoresis}. The balance of surface stress and Marangoni forces on the colloids defines their phoretic velocity in the radial direction $\hat{\textbf{e}}_r$ as \cite{wurger2007thermophoresis} 
\begin{equation}\label{eq:velocity}
\textbf{u}(r) = -\gamma_{T}\frac{\kappa R_{\rm TiO_2}}{3 \eta} \frac{\partial T(r)}{\partial r} \hat{\textbf{e}}_r,
\end{equation} 
where $\eta$ is the solvent (ethanol) viscosity, $\kappa = 3 \kappa_{\rm EtOH} / (2 \kappa_{\rm EtOH} + \kappa_{\rm TiO_2})$, with $\kappa_{\rm EtOH} = 0.171 \, {\rm W m^{-1} K^{-1}}$ and $\kappa_{\rm TiO_2} = 8.5 \, {\rm W m^{-1} K^{-1}}$ being the thermal conductivities of ethanol and ${\rm TiO_2}$ respectively, and $\gamma_T \approx -40 \, {\rm nJ \, m^{-2} \, K^{-1}}$ is the only fitting parameter with units of surface tension extracted from our data (Fig.~\ref{fig:Fig2}B). As $\gamma_T < 0$, particles migrate towards warmer regions (negative thermophoresis). Fig.~\ref{fig:Fig2}B and the arrows in Fig.~\ref{fig:Fig2}A show how the magnitude of $\textbf{u}(r)$ increases, both in experiments and model, as $\sim r^{-2}$ when nearing the heat source, so that, as shown by the simulated trajectory in Fig.~\ref{fig:Fig2}A \cite{Supplementary}, the motion of a single ${\rm TiO_2}$ particle becomes more directed with proximity to the source. The experimental dissipative dynamics of accumulation, dispersal and re-accumulation of the colloidal assembly can be thus well replicated with a simple two-dimensional particle-based model that includes this thermophoretic drive towards the heat source and a short range repulsive interaction among the colloids in the order of $\sim k_{\rm B} T$ (Fig.~\ref{fig:Fig2}C) \cite{Supplementary}. The cluster rate of growth is steeper at the start of the process as colloids within a $\sim 4 R_{\rm s}$ distance from the Janus particle experience stronger drifts ($\gtrsim 1 \, {\rm \mu \, s^{-1}}$, Fig. \ref{fig:Fig2}B). As close-by colloids are quickly drawn towards the heat source, those that are farther away take longer to arrive as their motion is dominated by their Brownian dynamics (less directed) due to $\textbf{u} \to \textbf{0}$ as $\partial T / \partial r \to 0$ (i.e. for $r \to \infty$). Eventually, this trend plateaus in time as colloids become depleted near the Janus particle and due to the finite extent over which the heat source induces significant flows \cite{makey2020universality}. When the heat source is turned off, the assembly dissolves driven by diffusion and a short-ranged interparticle repulsion with a rate of dispersal tending to zero as colloids evolve towards an equilibrium distribution (Fig.~\ref{fig:Fig2}C) \cite{Supplementary}. When the heat source is back on, colloids reaccumulate much faster than in the first accumulation phase due to the now higher colloidal density around the heat source (Fig.~\ref{fig:Fig2}C and \ref{fig:SI_Reaccumulation}).

Random lasing action occurs for a strongly enough scattering medium with a  short enough scattering mean free path $\ell_\mathrm {sc}$ and large enough excited cluster (as in Fig.~\ref{fig:Fig1}B). This  condition can be quantified by the critical radius $R_\mathrm{cr}$, function of $\ell_{\rm sc}$, as lasing action can only be achieved if the excited random laser area is of radius $R_\mathrm{ex} > R_\mathrm{cr}$ ~\cite{DeterminingRL}. For colloidal assemblies, $R_\mathrm{cr}$ can be calculated by solving the radiative-transfer equation for light \cite{Supplementary}. In our experiments, $R_\mathrm{cr}$ is a time-dependent variable as $\ell_{\rm sc}$ changes in time due to the accumulation and dispersal of colloids. Fig.~\ref{fig:Fig2}D reveals how $R_{\rm cr}$ decreases monotonically with increasing particle densities, thus facilitating the narrowing of the linewidth and lasing action. This trend is confirmed by the equivalent experimental data in Fig.~\ref{fig:Fig2}E, where we determine $R_\mathrm{cr}$ dependence on particle density by pumping large colloidal assemblies with different ${\rm TiO_2}$ concentrations at a fixed pump fluence (140 mJ cm$^{-2}$) while varying the pump spot size  \cite{Supplementary}. Based on Fig. \ref{fig:Fig1}E-F, $R_{\rm cr}$ is taken as the pump size where the linewidth of the emission spectrum narrows to half of its initial value, i.e. to 13.5~nm, in good agreement with model predictions (Fig.~\ref{fig:Fig2}D), where the only fitting parameter is the net gain length~\cite{Supplementary}. The particle densities before and after accumulation are marked on Fig.~\ref{fig:Fig2}D showing that $R_\mathrm{ex}$ is smaller (larger) than $R_\mathrm{cr}$ before (after) accumulation, thus explaining why we observe lasing only after a certain accumulation time, i.e. for large enough assemblies (larger than $R_\mathrm{cr}$), yet not before accumulation or after dissipation.

\begin{figure}[htb]
    \centering
    \includegraphics[width=12 cm]{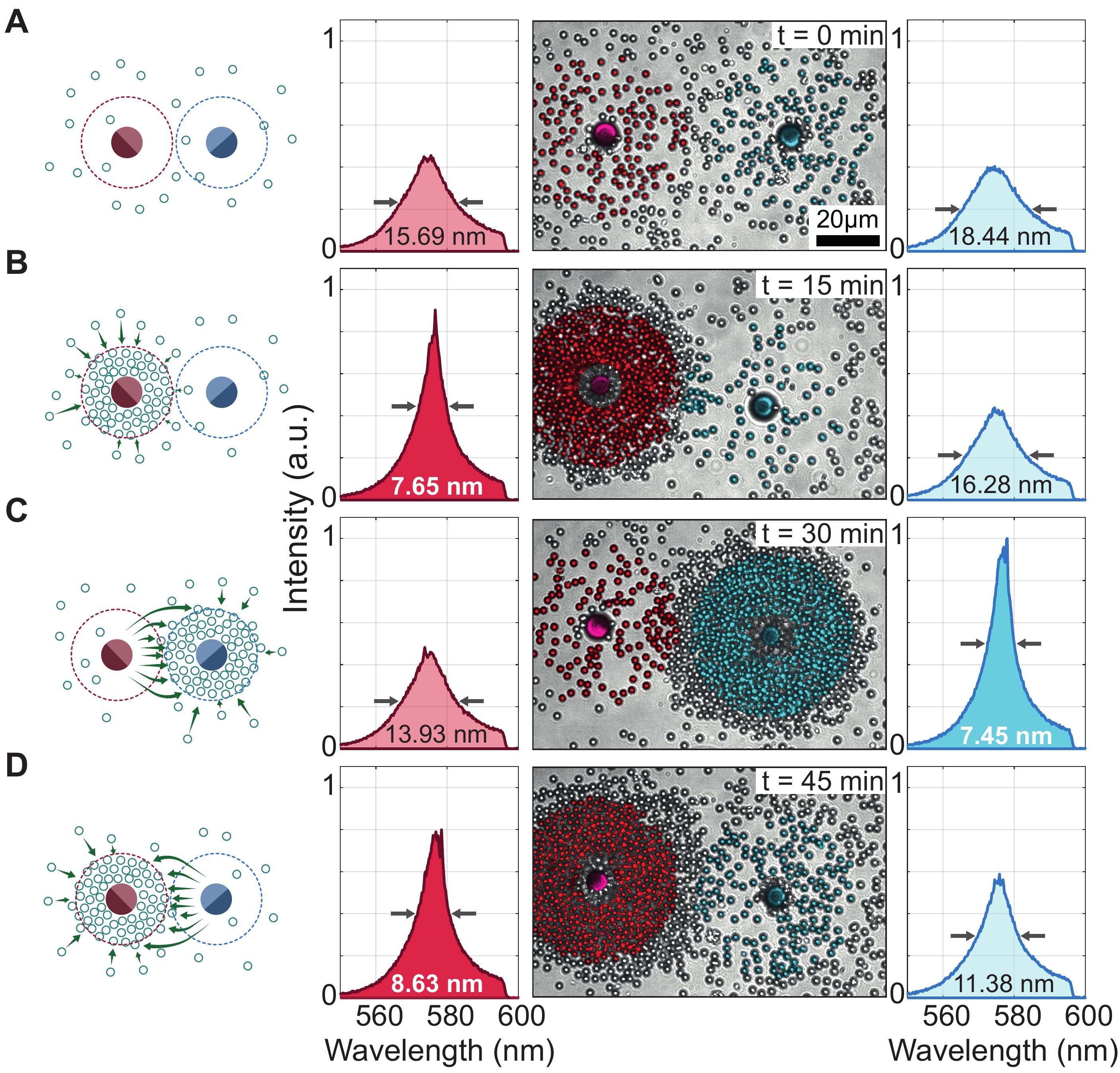}
    \caption{{\bf Reconfiguring random lasers by load transfer.} Two Janus particles (red and blue) are positioned $\sim$60 $\mu$m apart in a TiO$_2$ and laser dye colloidal solution. The region around each particle is pumped separately with the pump laser at a constant fluence (54 mJ cm$^{-2}$) for 15 minutes each time. The shaded surrounding colloids highlight the respective pump areas. The side panels show the measured emission spectra and respective linewidths at the end of each accumulation period. ({\bf A}) Before the initial accumulation, lasing is not observed from either region. ({\bf B}) When the left Janus particle is active as heat source, lasing is observed after accumulation. ({\bf C}) The colloidal load can then be transferred to the right Janus particle and ({\bf D}) back by alternating their use as heat sources, thus triggering selective lasing action in a different region at each transfer.
    }
    \label{fig:Fig3}
\end{figure}

The dynamic nature of the dissipative self-organization behind our random lasers, linking structure with optical functionality, can be harnessed to achieve unconventional tasks for standard random lasers, such as reconfigurability in space and time. Employing heat sources at different locations, lasing action can be triggered first and then transferred in space,  thus shifting the lasing load within the sample. Fig.~\ref{fig:Fig3} shows how two Janus particles, placed approximately one pump spot size apart, alternatively act as accumulation points for the ${\rm TiO_2}$ colloids, when respectively activated by the external energy source. Lasing action is then switched from one heat source to the other by transferring the colloidal scattering load across, and back. 

\begin{figure}[htb]
    \centering
    \includegraphics[width=10 cm]{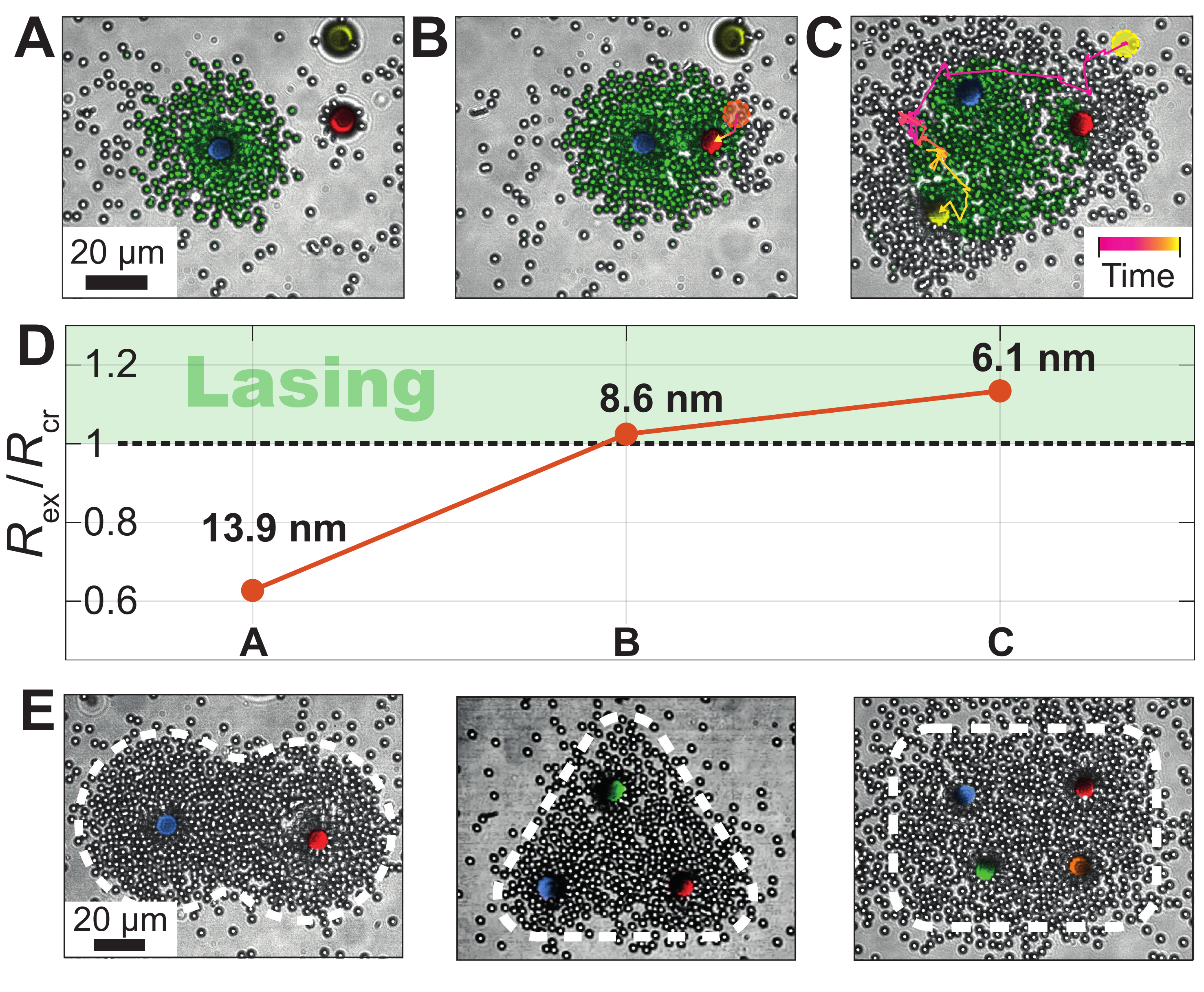}
    \caption{{\bf Cooperative random lasing.} ({\bf A-D}) Multiple Janus particles can act cooperatively to boost lasing action by fusing different colloidal clusters. ({\bf A}) A small cluster of ${\rm TiO_2}$ colloids around a single illuminated Janus particle (blue) does not reach the lasing threshold (dashed line in {\bf D}): it has a large linewidth of 13.9~nm and a very low excited system size, $R_\mathrm{ex}/R_\mathrm{cr}< 1$ (dots in {\bf D}). ({\bf B}) The cooperative action of a second Janus particle (red) drives the laser just above threshold, ({\bf C}) while the cluster formed with the cooperation of a third particle (yellow) shows an even better lasing action with a small linewidth (6.1~nm) and $R_\mathrm{ex}/R_\mathrm{cr} = 1.13$. The trajectories of the second and third Janus particles are highlighted in \textbf{B-C}. The green shaded area in {\bf A-C} shows the pump spot size of 52~$\rm \mu m$ at a fluence of 44 mJ cm$^{-2}$. ({\bf E}) Three examples of lasing cluster geometries (linewidths of 6.1 nm, 5.7 nm, and 4.7 nm, repectively) achievable when multiple Janus particles cooperate (a hourglass, a triangle, a rectangle), highlighting the structural flexibility of the colloidal assemblies. 
    }
    \label{fig:Fig4}
\end{figure}

Janus particles can also cooperate to achieve feats beyond what achievable by a single particle. Fig. \ref{fig:Fig4} shows how cooperation of properly located heat sources can lead to boosting the lasing properties and morph the laser spatially by imparting different shapes to the colloidal assemblies. While the colloidal assembly in Fig. \ref{fig:Fig4}A is too small to lase as $R_\mathrm{ex}<R_\mathrm{cr}$ (Fig. \ref{fig:Fig4}D), the addition of further colloids by a second Janus particle (Fig. \ref{fig:Fig4}B) pushes the cluster size just above the threshold for lasing ($R_\mathrm{ex}\approx R_\mathrm{cr}$ in Fig. \ref{fig:Fig4}D). This effect becomes even more pronounced when a third Janus particle joins the assembly as $R_\mathrm{ex}> R_\mathrm{cr}$ (Fig. \ref{fig:Fig4}C-D). Beyond the possibility of refining the laser spectral properties, another alluring application for multiple cooperating Janus particles is the possibility to define different planar laser geometries (Fig. \ref{fig:Fig4}E), which could be important for display applications, for traveling through narrow channels, or for adhering to complexly shaped targets. 


In conclusion, we have performed the first experimental demonstration of an artificial colloidal material that, by virtue of dissipative self-assembly dynamics, can spontaneously self-organize in a random laser device, which dynamically blends morphology with its optical functionality. These self-organized random lasers can be manipulated on demand to produce controllable and programmable lasing, thus bringing novel functionalities to the field of photonics and opening the door for a new class of active functional materials in active matter. 
Their life-like features (responsiveness, reconfigurability and  cooperation) are a first step towards the realization of fully animate lasers capable of independent motion and autonomous adaptation in response to external stimuli \cite{AnimateMaterials}, thus helping narrow the gap between living and artificial self-assembled materials.
We envisage that the realization of similar self-organized lasers from light-actuated colloidal molecules \cite{singh2017non, Schmidt2019} will pave the way towards the development of a new class of bio-inspired functional materials with potential for sensing applications~\cite{GaioSensing}, non-conventional computing \cite{phillips2014digital} and novel light sources and displays \cite{kim2011self}. 

\begin{bibunit}[plain]
    \bibliography{scibib}
    \bibliographystyle{Science}
\putbib[mybib]
\end{bibunit}

\paragraph*{Acknowledgments.} We are grateful to Samantha Rueber, Matthew Blunt and Valentino Barbieri for initial training on experimental techniques. {\bf Funding:} GV acknowledges sponsorship for this work by the US Office of Naval Research Global (Award No. N62909-18-1-2170). WKN acknowledges the research support funded by the President's PhD Scholarships from Imperial College London. RS and DS acknowledges support from The Engineering and Physical Sciences Research Council (EPSRC), grant number EP/T027258, and the European Community. {\bf Author contributions:} Author contributions are defined based on the CRediT (Contributor Roles Taxonomy) and listed alphabetically. Conceptualization: RS, GV. Data Curation: WKN, MT. Formal analysis: WKN, DS, MT, GV. Funding acquisition: RS, GV. Investigation: WKN, RS, DS, MT, GV. Methodology: WKN, RS, MT, GV. Project administration: RS, DS, MT, GV. Software: WKN, DS, GV. Supervision: RS, GV. Validation: WKN, DS, MT. Visualization: All. Writing – original draft: All. Writing – review and editing: All. {\bf Competing interests:} The authors have no competing interests. {\bf Data and materials availability.} All data in support of this work is available in the manuscript or the supplementary materials. Further data, codes or materials are available from the corresponding authors upon reasonable request.

\paragraph*{Supplementary Materials} \

Materials and Methods

Fig \ref{fig:SI_TiO2Diffusion} – \ref{fig:SI_NoLasing}

References \cite{tang2006effect, Crocker1996, Warrick1992, Branka1999, AJPVolpe2015, Guerin2016, Holzer2000}

\clearpage
\setcounter{page}{0}
\pagenumbering{arabic}
\setcounter{page}{1}

\section*{Supplementary Materials}

\subsection*{Materials and Methods}

\paragraph*{Materials.}

Glass microscopy slides (Thermo Fisher) were purchased from VWR while glass coverslips were purchased from Thorlabs. The following chemicals were purchased and used as received: rhodamine 6G (Sigma-Aldrich), rhodamine B (Acros Organics), acetone ($\ge$ 99.8$\%$, Sigma-Aldrich), ethanol ($\ge$ 99.8$\%$, Fisher Scientific), sodium hydroxide (NaOH, Fisher Scientific), polyethylenimine (PEI, branched, Mw ~25,000, Mn ~10,000, Sigma-Aldrich). Deionized (DI) water ($\ge$ 18 M$\Omega.$cm) was collected from a Milli-Q purification system. Aqueous colloidal dispersions (5$\%$ w/v) of silica (${\rm SiO_2}$) colloids for fabricating Janus particles and to be used as spacers (8.44 $\pm$ 0.27 $\mu$m and 20 ± 0.64 $\mu$m in diameter, respectively) were purchased from Microparticles GmbH. Aqueous colloidal dispersions of titania (${\rm TiO_2}$) particles (2.5$\%$ w/v, 1.73 ± 0.03 $\mu$m in diameter) for the lasing experiments, of fluorescent ${\rm SiO_2}$ particles (2.5$\%$ w/v, 10.05 $\pm$ 0.31 $\mu$m, excitation/emission: 602 nm/623 nm) for Fig. \ref{fig:SI_JanusOrientation} and of polystyrene particles (10$\%$ w/v, 1.65 $\pm$ 0.04 $\mu$m) for Fig. \ref{fig:SI_NoLasing} were also purchased from Microparticles GmbH. Carbon rods of length 300 mm and diameter 6.15 mm for coating Janus particles were purchased from Agar Scientific and cut to a length of 50 mm before use. UV cure adhesive (Blufixx) and hydrophobic coating (RainX) for sample preparation were purchased from an online retailer (Amazon). 

\paragraph*{Slide cleaning protocol.} 
Before their use for sample preparation, glass slides and coverslips were cleaned via sonication for 10 min in 2 M NaOH ethanolic solution followed by three cycles of 5 min sonication in DI water. To dry them, the slides were withdrawn from the water in the presence of ethanol vapor (Marangoni drying) and, subsequently, blown with a nitrogen gun.

\paragraph*{Fabrication of Janus particles.} 
The Janus particles used in our experiments as heat sources were fabricated from ${\rm SiO_2}$ colloids of radius $R_{\mathrm s} = 4.22 \pm 0.14 \, {\rm \mu m}$, which were coated on one side with a thin layer ($\approx60 \, {\rm nm}$) of carbon. We first deposited a monolayer of colloids on a clean glass slide. The monolayer was obtained by evaporating a $40 \, {\rm \mu L}$ droplet containing a $2.5 \% \, {\mathrm w}/{\mathrm v}$ dispersion of the colloids in DI water. The particles were then coated with a 60 nm thick carbon layer using an automatic carbon coater (AGB7367A, Agar Scientific). Post-coating sonication allowed us to dislodge the half-coated particles in DI water from the glass slides to use them for sample preparation.

\paragraph*{Preparation of samples of colloids and laser dyes.} 

Random lasers combine optical gain with a scattering medium~\cite{DeterminingRL}. The samples used are in the form of a glass chamber (see next section) filled with a colloidal dispersion in an ethanol solution of laser dye. The optical gain is provided by two types of rhodamine-based dyes in ethanol solution ($1\%$ w/v): rhodamine 6G (Rh6G) for Figs. \ref{fig:Fig1}, \ref{fig:Fig2}, \ref{fig:SI_Reaccumulation} and \ref{fig:SI_NoLasing}, and rhodamine B (RhB) for Figs. \ref{fig:Fig3}, \ref{fig:Fig4} and \ref{fig:SI_RhB_Rcr}. The use of the two dyes is motivated by their complementary features in experiments: at $\lambda = 532$~nm (the wavelength of the pump laser), pumping rhodamine 6G (absorption peak at $530$~nm) is more efficient than pumping rhodamine B (absorption peak at $\lambda = 550$~nm), thus improving lasing performance at a given power (i.e. narrower linewidth); RhB solutions instead reduce sticking of Janus particles to the glass substrate, thus performing better for tasks where their maneuverability is paramount. The scattering medium is a dispersion of monodisperse ${\rm TiO_2}$ colloids, chosen due to the material's characteristic high refractive index ($n_{\rm TiO_2} \approx 2.3$ for amorphous titania) larger than that of ethanol ($n_{\mathrm EtOH} = 1.36$). This refractive index difference is important to achieve the strong scattering properties needed for lasing action: in fact, no lasing is observed when substituting the ${\rm TiO_2}$ colloids with lower refractive index polymer colloids of similar size (Fig. \ref{fig:SI_NoLasing})~\cite{PhotonicGlassRL}. 
We obtained stable dispersions of ${\rm TiO_2}$ colloids in 1\% w/v ethanol solutions of laser dyes (either rhodamine 6G  or rhodamine B) by functionalizing the colloids with PEI to prevent flocculation and sticking to the glass substrate \cite{tang2006effect}. In particular, we first mixed $50 \, {\rm \mu L}$ of a 0.1\% aqueous dispersion of Janus particles with stock solutions of ${\rm TiO_2}$ colloids ($20 \, {\rm \mu L}$) and $20 \, {\rm \mu m}$ ${\rm SiO_2}$ colloids ($10 \, {\rm \mu L}$) in a 1.5 mL centrifuge tube (Eppendorf). The $20 \, {\rm \mu m}$ ${\rm SiO_2}$ colloids were added to act as spacers in the sample chambers. This colloidal cocktail is then centrifuged at 1000 RCF for 3 min leaving a pellet; the supernatant is removed and the pellet is redispersed in $100 \, {\rm \mu L}$ of a 3\% w/v ethanol solution of PEI. The dispersion is then sonicated for 5 min to fully redisperse the ${\rm TiO_2}$ colloids and to allow for functionalizing their surface with PEI, as confirmed by Fourier-transform infrared spectroscopy (FTIR) in Fig. \ref{fig:SI_Functionalization}. After functionalization with PEI, the size of the ${\rm TiO_2}$ colloids increased to 1.83 ± 0.05 $\mu$m from 1.73 ± 0.03 $\mu$m of the pristine particles as confirmed by scanning electron microscopy measurements. The dispersion is left to rest for 30 min to make sure all particles are sufficiently coated with the polymer. Finally, it is centrifuged again at 1000 RCF for 2 min, the supernatant removed and replaced with $50 \, {\rm \mu L}$ of a 1\% w/v ethanol solution of a rhodamine dye.

\paragraph*{Sample chamber preparation.} Experiment-ready sample chambers containing a dispersion of colloids in ethanol solutions of laser dyes were prepared by sandwiching $15 \, {\rm \mu L}$ of the dispersion between a clean glass slide and a thin coverslip using low concentrations of $20 \, {\rm \mu m}$ silica particles as spacers. Prior to this, both slide and coverslip were soaked for 2 min in Rain-X, a commercial solution which renders glass surfaces more hydrophobic and aids limiting particle sticking to the glass chamber. Excess RainX was removed by soaking the slide in acetone and subsequently wiping with lens tissue. The chamber was then sealed by applying a UV curable adhesive around the borders of the coverslip, taking care of not exposing the dye solution to UV light by illuminating only the edges of the coverslip, as this could cause dye bleaching. Before data acquisition, the sample was left to equilibrate over a one-hour period.

\paragraph*{Optical setup and microscopy.} 
Fig. \ref{fig:SI_Setup} shows a schematic of the experimental setup used to illuminate the Janus particles, to image the sample and to probe its emission spectra. Samples are mounted on the stage of a Nikon Ti microscope. Two laser sources are exploited: a continuous-wave HeNe laser (Thorlabs, $\lambda=632.8$~nm, CW, $20$~mW) and a Nd:YAG pulsed laser (TEEM Power-Chip, $\lambda=532$~nm, pulse width 400~ps, 20~$\mu$J energy per pulse, 1-1000 Hz). The HeNe laser is used as a energy source for heating the Janus particles, while the pulsed laser is used to reach population inversion for lasing measurements. The choice of using the HeNe as energy source was dictated by the need to avoid overlap with the absorption spectrum of rhodamine dyes (centered at 525 nm), thus avoiding spurious heating effects in the sample due to this competing absorption process in the dye. The pulsed laser is only operating when measuring emission spectra and in single shot mode for less than 30 s each time to limit absorption from the Janus particle and to avoid modification of the accumulated colloidal cluster.

The HeNe laser is coupled to a single-mode optical fiber and the beam from the fiber is focused onto the sample with a lens of 60 mm focal length to a spot diameter of $\sim5~\mu$m. This spot size was chosen to match the size of the Janus particle and for ease of alignment while making sure the Janus particle is heated with sufficient power density. For accumulation of ${\rm TiO_2}$ colloids around the Janus particle to occur, the laser power density at the sample was around 0.14~mW $\rm \mu m^{-2}$. The laser power was controlled by a movable knife edge before the fiber-coupler. The same HeNe laser was also used to manipulate the Janus particles by optical forces or cavitation (see next section).

For lasing measurements, samples were optically pumped at room-temperature with the Nd:YAG pump laser. The pump laser profile was shaped with a programmable digital micro-mirror device (DMD, Ajile AJD-4500), and the excitation pattern was imaged onto the sample through a 40$\times$ objective lens (Nikon CFI Plan Fluor 40X, 0.75 N.A., 0.66 mm W.D.). An acousto-optic modulator (AOM) is used to control the energy of the pump laser. Circular illumination profiles, of constant intensity in a disk shape, and radius $R_\mathbf{ex}$, were used for the measurements in Figs. \ref{fig:Fig1}, \ref{fig:Fig2}, \ref{fig:Fig4}, \ref{fig:SI_Reaccumulation} and \ref{fig:SI_RhB_Rcr}. Doughnut-shaped profiles, of constant intensity in a disk  shape without the central part, centered on the Janus particle were used for Fig. \ref{fig:Fig3} and \ref{fig:SI_NoLasing} to avoid further laser exposure and prevent any motion of the Janus particle. The size of the missing central part was chosen to be about 1.5 times bigger than the size of the Janus particle for ease of alignment. When the sample was pumped, the HeNe laser was blocked to avoid overheating the Janus particle, and in particular to minimize the formation of cavitation bubbles during the lasing measurements. The lasing emission from the sample was collected through the same objective lens, filtered, and then focused into a stripe on the spectrometer entrance slit via a cylindrical lens to maximize the measured signal counts. The signal was spectrally analyzed using a grating spectrometer (Princeton Instruments Isoplane-320) equipped with a 600~gr mm$^{-1}$ visible grating ($0.5$~nm resolution) and a CCD camera (Princeton Instruments Pixis 400). 
Linewidths are given as full width at half maximum (FWHM).

The critical radius $R_\mathbf{cr}$ in Fig.~\ref{fig:Fig2} was measured at constant pump fluence of 140 mJ cm$^{-2}$, by changing the size of the illumination spot of radius $R_\mathbf{ex}$ while recording the emission spectrum, until the emission linewidth reached the value of 13.5~nm, which is half the fluorescence linewidth measured at low pumping powers.

The motion of the ${\rm TiO_2}$ colloids in Figs. \ref{fig:Fig1}, \ref{fig:Fig2} and \ref{fig:SI_Reaccumulation} was recorded using a CMOS camera (Thorlabs) at a frame rate of 2 fps (frames per second). Image focus was adjusted so that each particle had a bright spot at its center relative to the background to provide enough contrast to discern individual particles via digital video microscopy based on a homemade MATLAB tracking software \cite{Crocker1996}.

\paragraph*{Manipulation of Janus particle.} In order to position the Janus particles at different locations in the sample chamber, we manipulated them with the mildly focused HeNe laser, which, at power densities of  0.14~mW $\rm \mu m^{-2}$, exerts a gentle pulling optical force which predominantly drags the Janus particle towards the center of the laser spot. The manipulation was performed by either moving the sample stage only (Figs. \ref{fig:Fig1}, \ref{fig:Fig2}, \ref{fig:Fig3}, \ref{fig:SI_Reaccumulation} and \ref{fig:SI_NoLasing}) or the HeNe laser spot only (Fig. \ref{fig:Fig4}), while keeping the other element fixed. The same HeNe laser at higher power densities (above  0.2~mW $\rm \mu m^{-2}$) was also used to free the Janus particles in $\rm TiO_{2}$ clusters by cavitation before repositioning them (Figs. \ref{fig:Fig3} and \ref{fig:Fig4}). Cavitation is created when the Janus particle is strongly heated by the HeNe laser leading to bubble formation (Fig. \ref{fig:SI_TMeasurement}) and provides an instantaneous pushing action against the carbon-side of the Janus particle that propels it away from the surrounding $\rm TiO_{2}$ colloids. 

\paragraph*{Calculation of the temperature profile around a heat source.} 
As the dynamics of our dissipative colloidal assemblies around an illuminated Janus particle predominantly take place near the glass surface, we can calculate the temperature profile generated by a disc heat source of radius $R_{\mathrm s}$ at temperature $T_{\mathrm s}$ in two dimensions. This is a reasonable approximation considering that the most likely configuration for the Janus particle in a formed cluster is with the cap facing down, i.e. towards the interface (Fig. \ref{fig:SI_JanusOrientation}). Under continuous illumination, the heating of the surrounding fluid can be assumed instantaneous so that a steady-state temperature profile around the heat source is promptly reached. In fact, since heat propagation is much faster than particle migration, the temperature field can be considered as stationary. Assuming a steady state for the diffusion of heat from the source, we can then calculate the temperature profile around the heat source in the plane of motion as \cite{Warrick1992}:

\begin{equation}\label{T_profile}
	T(r) = \frac{2}{\pi} (T_{\rm s} - T_{\rm b}) \sin^{-1}\bigg( \frac{R_{\mathrm s}}{r} \bigg) + T_{\mathrm b}, \, \mathrm{for} \, r \ge R_{\mathrm s}
\end{equation}
where $r$ is the radial distance from the center of the heat source and $T_{\rm b}$ is room temperature. Note that, in steady state, the presence of an interface does not influence the temperature profile. 

\paragraph*{Particle-based simulations.} 
We consider a simple numerical model where $N$ hard spheres of mean radius $R_{\rm TiO_2}$ move inside a two-dimensional square box of side $B =  187 \, {\rm \mu m}$. Particles were placed at random without overlap at fixed density (0.05 particles ${\rm \mu m^{-2}}$). Otherwise specified differently, the values for all parameters in the simulations are set to the exact experimental values reported in the main text. The radius of each individual particle $R_i$ is taken from a Gaussian distribution with mean $R_{\rm TiO_2}$ and standard deviation $\delta R$ to reproduce the size variability of the monodisperse sample of ${\rm TiO_2}$ colloids. At the center of the box, we place a disc heat source of radius $R_{\mathrm s}$. When the source is active, it has a uniform temperature $T_{\mathrm s}$ and generates a radial temperature profile according to Eq. \ref{T_profile}. When the source is not active, the temperature is uniform in the box and equivalent to room temperature $T_{\mathrm b}$.

The trajectory of the $i$-th particle is then obtained by solving the following Langevin equation in the overdamped regime using the second-order stochastic Runge-Kutta numerical scheme \cite{Branka1999}

\begin{equation}\label{eq:ModelLangevin}
\dot{\mathbf{x}}_i = \mathbf{u}_i + \sum_{j\ne i}{\frac{\mathbf{F}_{ji}}{\gamma_i}} + \frac{\mathbf{F}_{{\mathrm s}i}}{\gamma_i} + \sqrt{2 D_i}\bm{\xi}_i,
\end{equation}
where $\mathbf{x}_i$ and $\mathbf{u}_i$ are respectively the particle's position and thermophoretic velocity at time $t$ (Eq. \ref{eq:velocity}), $D_i = k_{\rm B}T_i/\gamma_i$ is the particle's diffusion coefficient at position $\mathbf{x}_i$ and temperature $T_i$ (Eq. \ref{T_profile}) with $\gamma_i$ its friction coefficient \cite{AJPVolpe2015}; $\bm{\xi}_i$ is a two-dimensional vector of independent white noise process with zero mean and unitary variance \cite{AJPVolpe2015}. When the heat source is off, $\mathbf{u}_i$ is null. The direction of motion due to the thermophoretic velocity is therefore defined by the unitary vector $\hat{\mathbf{e}}^i_r(t) = \left [ \cos(\theta_i(t)), \, \sin(\theta_i(t) \right ]$, where $\theta_i(t)$ is the particle's angular coordinate in the frame of reference defined by the heat source.

We implemented particle-particle steric interactions $\mathbf{F}_{ji}$ with the repulsive term of a Lennard-Jones potential with parameters $\epsilon = k_{\rm B} T_{\rm b}$ and $\sigma = 2R_{\rm TiO_2}$. The effect of this repulsive term is short ranged and was truncated at $1.6(R_i+R_j)$ for each neighbouring particles $i$ and $j$. Finally, we modelled the steric interaction with the heat source by introducing a repulsive force $\mathbf{F}_{{\mathrm s}i}(r_i)$ in the equation of motion. This force depends on the particle's distance $\mathbf{r}_i$ from the heat source as

\begin{equation}\label{eq:ModelForce}
\mathbf{F}_{{\mathrm s}i}(r_i) \propto \frac{e^{-r_i}} {|r_i-R_s+R_i|} \hat{{\bf e}}^i_r. 
\end{equation}
This function was chosen to reproduce a  strong (local) repulsive interaction between particle and source, i.e. to mimic a hardcore potential. The exponential term ensures that the force does not increase too abruptly when approaching the source. This effect of this force was truncated at a cut-off distance $r_{\rm c} = 2.5R_{\rm s}$. 

\paragraph*{Random lasing model.} 
The random lasing model is based on the radiative transfer equation (RTE)~\cite{Guerin2016} and estimates the critical radius $R_{\rm cr}$, which is the minimum size of a colloidal cluster needed to achieve the lasing threshold. Since the lateral dimensions of our sample are much larger than its thickness, we used a two-dimensional model where $R_{\rm cr}$ is given by solving 
\begin{equation}\label{eq:ModelRTE}
\frac{J_{0}(\sqrt{2g(\ell_{\rm sc}^{-1} - g)} R_{\rm cr})}{J_{1}(\sqrt{2g(\ell_{\rm sc}^{-1} - g)} R_{\rm cr})} = \frac{\pi}{2} \frac{g}{\sqrt{2g(\ell_{\rm sc}^{-1} - g)}},
\end{equation}
where $J_{0}$ and $J_{1}$ are the Bessel functions of the first kind, $\ell_{\rm sc} = 1/(\sigma_{\rm sc} \rho)$ is the scattering length ($\sigma_{\rm sc}$ is the scattering cross-section and $\rho$ is the particle density), and $g = 1/\ell_{\rm g,dye}$ is the gain coefficient of the gain medium, being $\ell_{\rm g,dye}$ the dye gain length. 
In our self-organized lasers, $\ell_{\rm sc}$ and hence $R_{\rm cr}$ depend on the time-varying particle density in the accumulation/dispersal processes.

In order to find the critical radii of the clusters with different particle densities at different times, Eq. \ref{eq:ModelRTE} was solved numerically, with  $g$ a fitting parameter and $\ell_{\rm sc}$ obtained from Mie theory for ${\rm TiO_2}$ colloids of $1.83~\mu$m diameter and a refractive index $n_{\rm TiO_2} = 2.3$. We chose to fit $g$ as its exact value depends on off-plane scattering losses (i.e. when light escapes from the sample plane) and on minor absorption losses from the partially pumped dye and the Janus particle, and therefore it is usually larger than $\ell_{\rm g,dye}$. A good fit to the experimental data in Fig. \ref{fig:Fig2}D is obtained with $g_{\rm net} = 0.0721 \, \mu$m$^{-1}$ (corresponding to a net dye gain length $\ell_{\rm g, net} = 13.87 \, \mu$m). This gain length estimate is around 3.5 times larger than $\ell_{\rm g,dye} = 4~\mu$m given that, for rhodamine, $\rm \sigma_{g,dye} = 2\times10^{-20}~m^{2}$ \cite{Holzer2000} and that the dye density $\rm \rho_{\rm dye} = 1.257\times10^{25} \, molecules \, m^{-3}$  in a 1\% w/v ethanol solution (the molar mass of rhodamine is 479 g mol$^{-1}$).

The value of $\ell_{\rm sc}$ (as estimated from Mie scattering calculations and particle counting from bright-field images) reaches a minimum value of 13$~\mu$m for a dense cluster, much larger than the light wavelength, while for a dilute cluster is of the order of 100$~\mu$m. The optical thickness, i.e. the cluster diameter divided by $\ell_{\rm sc}$, is smaller than 6. This justifies our modeling of the random laser in the intermediate regime between the diffusive and ballistic limits~\cite{Guerin2016}.

\newpage

\renewcommand{\thefigure}{S\arabic{figure}}
\setcounter{figure}{0}

\subsection*{Supplementary Figures}

\begin{figure}[htb!]
    \centering
    \includegraphics[width=0.6\textwidth]{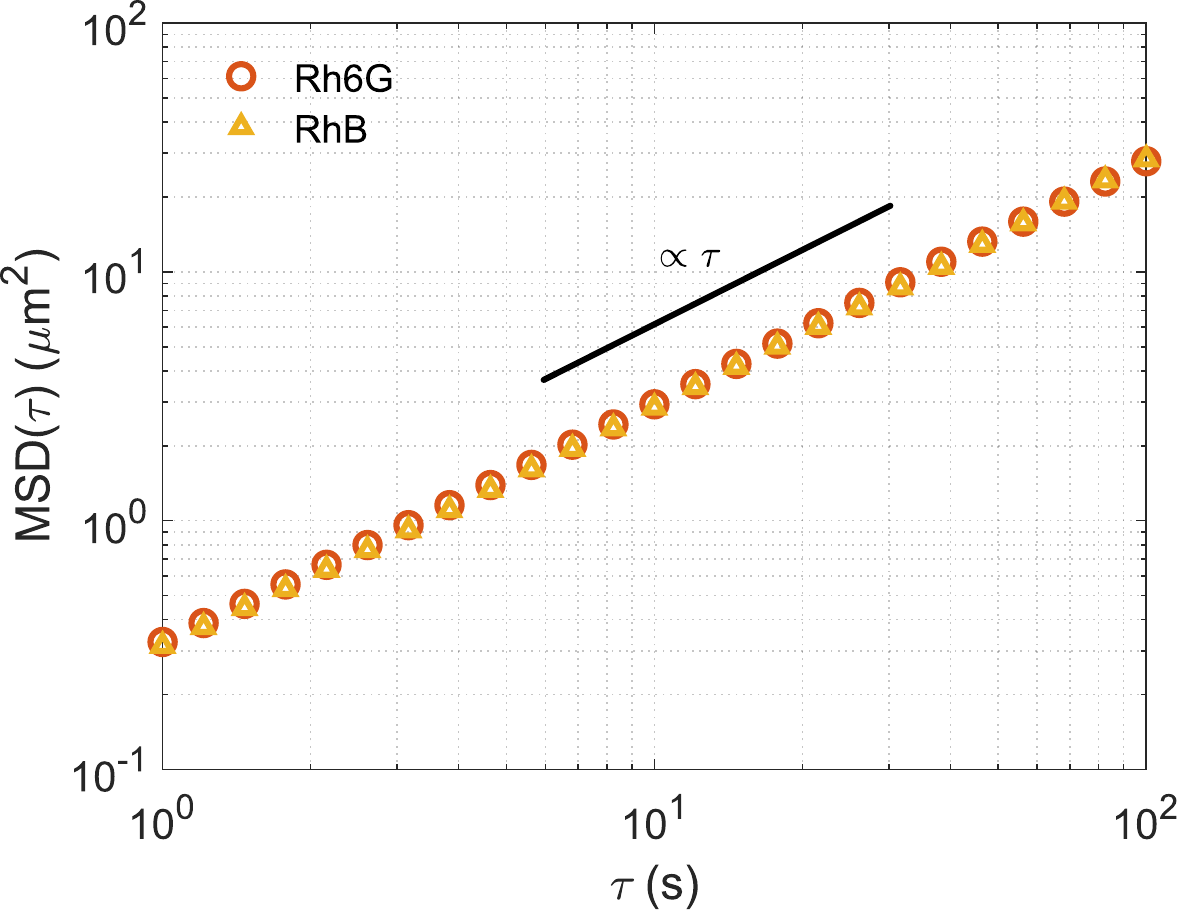}
    \caption{{\bf Diffusion of ${\rm \mathbf{TiO_2}}$ colloids in ethanol solutions of rhodamine dyes.} In our experiments, due to their size and non-negligible density ($\rho_{\rm TiO_2} = 4.24 \, {\rm g \, cm^{-3}}$), ${\rm TiO_2}$ colloids quickly sediment at the bottom surface of the experimental chamber, where, in the absence of a heat source, they primarily diffuse in 1\% w/v ethanol solutions of rhodamine dyes near the plane defined by this interface. The diffusive behavior is confirmed by the ensemble-averaged mean squared displacements (${\rm MSD}(\tau) \propto \tau$) for both rhodamine 6G (circles) and rhodamine B (triangles). Each MSD is calculated from the trajectories of at least 6 colloids. The experimentally measured diffusion coefficients are $D = 0.071 \pm 0.002 \, {\rm \mu m^2 \, s^{-1}}$ for colloids in rhodamine 6G and $D = 0.069 \pm 0.001 \, {\rm \mu m^2 \, s^{-1}}$ for colloids in rhodamine B. 
    } 
    \label{fig:SI_TiO2Diffusion}
\end{figure}

\begin{figure}[htb!]
    \centering
     \includegraphics[width=0.6\textwidth]{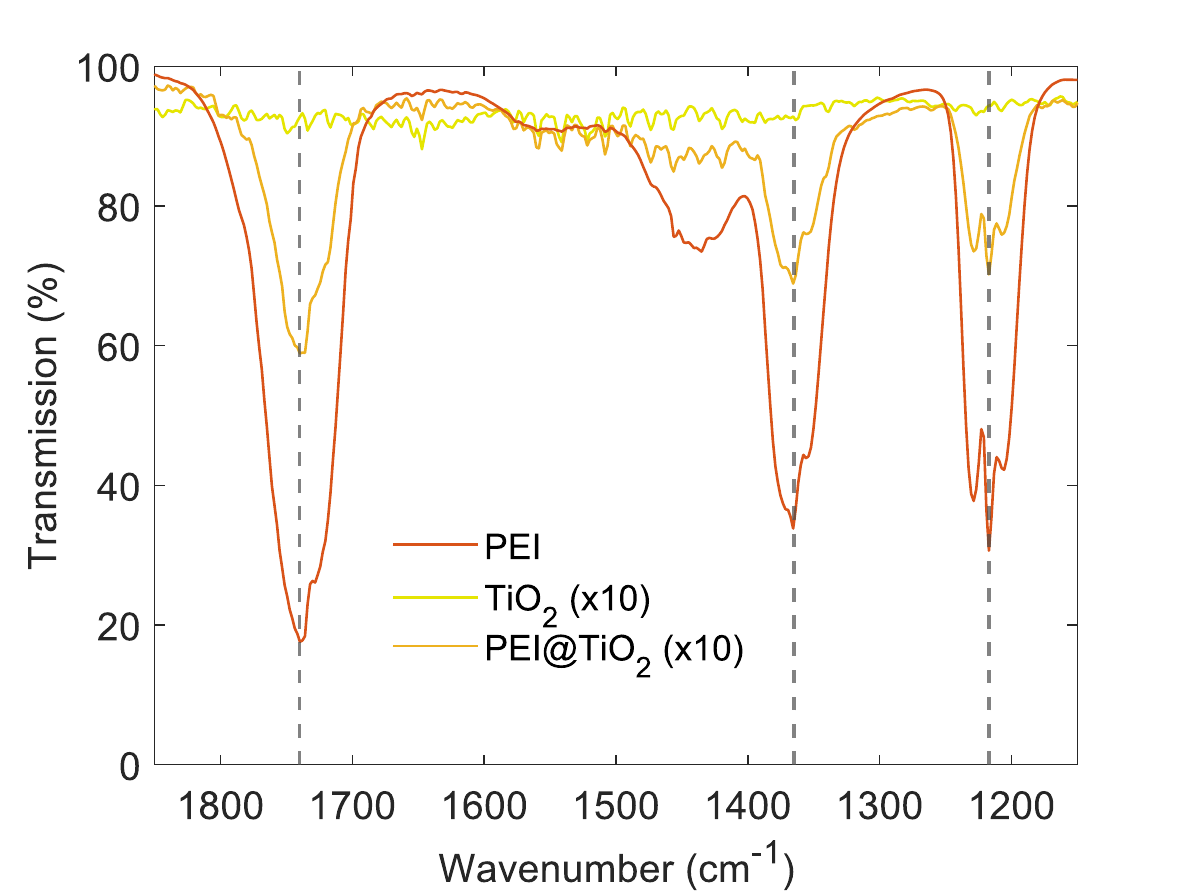}
    \caption{{\bf Functionalization of ${\rm \mathbf{TiO_2}}$ colloids}. Fourier-transform infrared spectroscopy (FTIR) measurements of polyethylenimine polymer (PEI), pristine titania colloids (${\rm TiO_2}$), and PEI-functionalized titania colloids (PEI${\rm@TiO_2}$). Notable peaks at 1740 $\rm cm^{-1}$, 1365 $\rm cm^{-1}$ and 1217 $\rm cm^{-1}$ (dashed lines) corresponding to the presence of PEI are found on the functionalized ${\rm TiO_2}$ colloids, but not on the pristine particles. Spectra (except that of PEI) are enhanced by a factor of 10 for an improved comparison.
    }
    \label{fig:SI_Functionalization}
\end{figure}

\begin{figure}[htb!]
    \centering
    \includegraphics[width=1\textwidth]{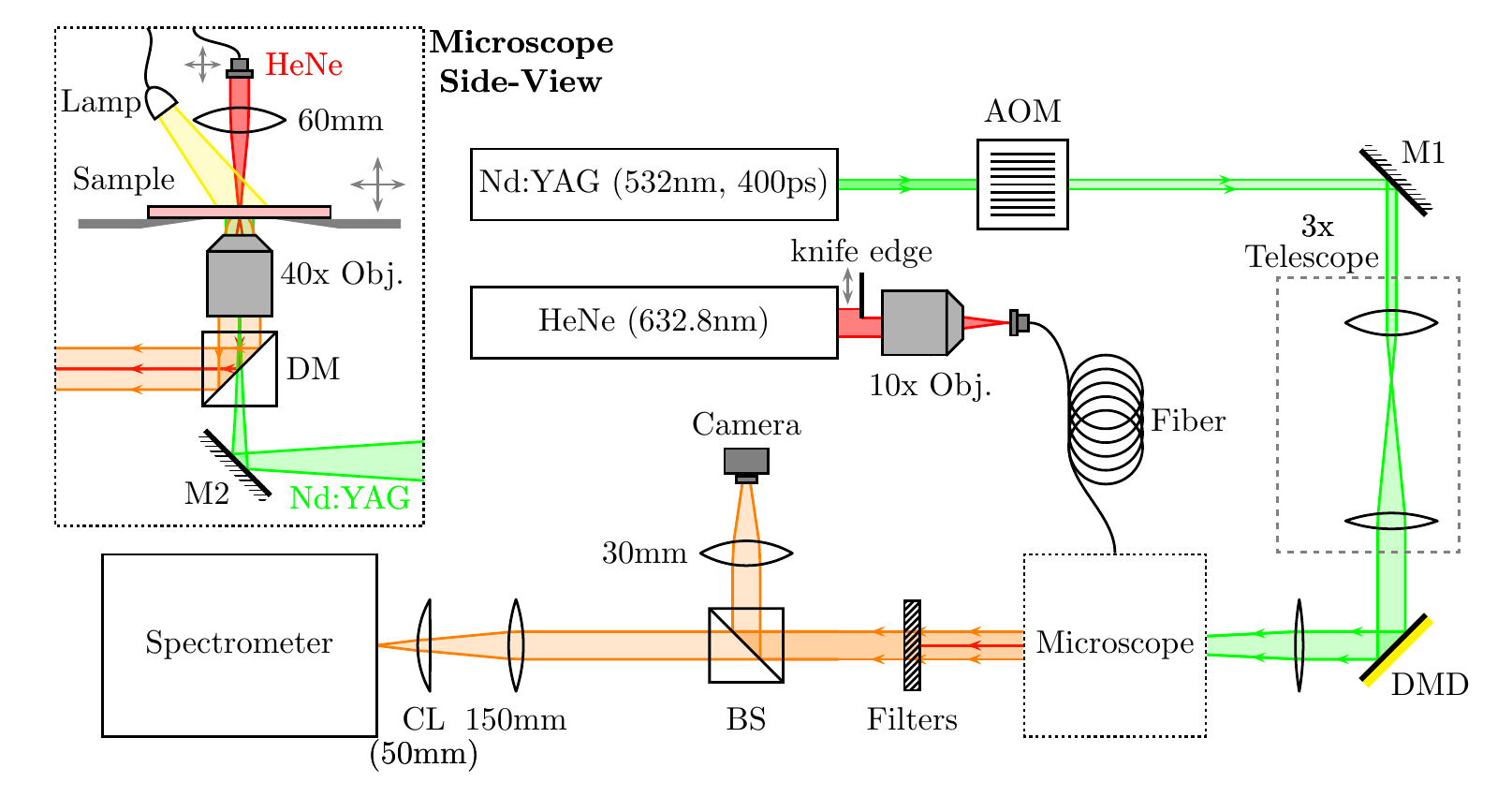}
    \caption{{\bf Experimental setup.} Two lasers (Nd:YAG and HeNe) are incorporated into a microscope where the sample with the colloidal particles is held (inset). The HeNe laser is used as an illumination source for the Janus particles, while the pulsed laser is used for lasing measurements. The pump power is controlled by an acoustic optical modulator (AOM), and a digital micro-mirror device (DMD) in the optical path is used to shape its excitation spot. The fibre-coupled HeNe laser is mounted on a separate stage and placed at the top of the sample. The lasing emission from the sample is collected and filtered by a 532~nm dichroic mirror (DM) and two filters (532~nm long-pass and 600~nm short-pass), then spectrally analyzed by a spectrometer. The lamp shown in the inset illuminates the sample for bright-field imaging. The sample image is then sent to a camera through a 30:70 beam splitter (BS). M: mirror; CL: cylindrical lens.
    }
    \label{fig:SI_Setup}
\end{figure}

\begin{figure}[htb!]
    \centering
    \includegraphics[width=0.6\textwidth]{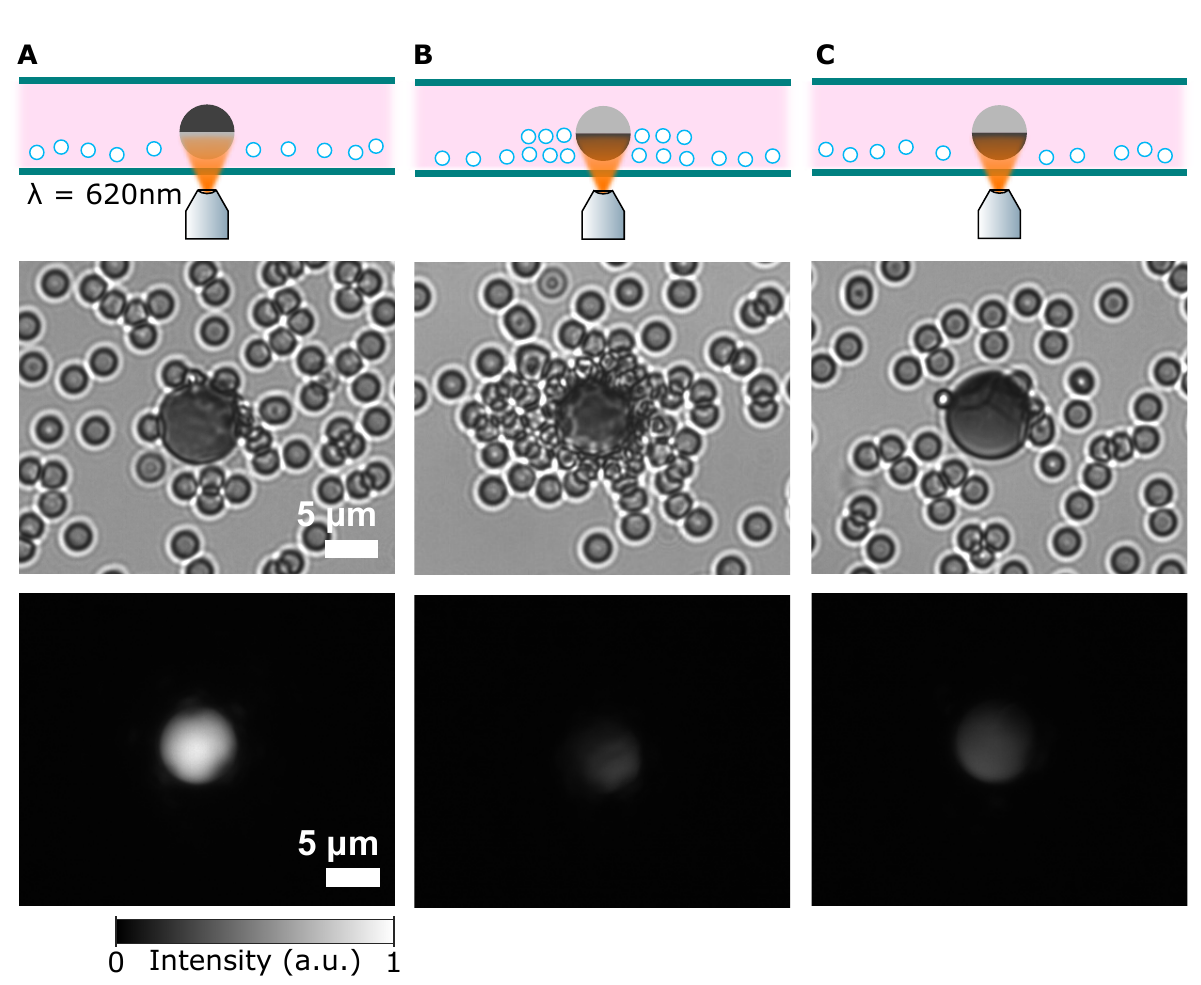}
    \caption{{\bf Orientation of Janus particles within colloidal assemblies.} Janus particles fabricated from fluorescent ${\rm SiO_2}$ colloids ($\approx 10 {\rm \mu m}$ in diameter, excitation peak at 602 nm; emission peak at 623 nm) were used to confirm the orientation of the Janus particles in our experiments after accumulation of ${\rm TiO_2}$ colloids. {\bf A-C} show (top) schematics and representative (middle) bright-field and (bottom) fluorescence images of the orientation of Janus particles under different conditions. ({\bf A}) A Janus particle is initially found in a cap-up configuration near the bottom surface of the sample chamber. When excited at 620~nm from the same side of the image detection, it appears bright under fluorescence imaging. ({\bf B}) After illuminating the particle in {\bf A} with a continuous-wave laser at 532 nm 
    to induce significant accumulation of ${\rm TiO_2}$ colloids, the particle is found to assume a cap-down configuration. In fact, when its translational motion is hampered by the surrounding cluster, the particle's diffusive rotational dynamics lead to a cap-down equilibrium orientation dictated by gravity, due to the higher density of carbon ($\rho_{\rm C} = 3.52 \, {\rm g \, cm^{-3}}$) with respect to silica ($\rho_\mathrm{SiO_2} = 2.65 \, {\rm g \, cm^{-3}}$). After having switched off the external illumination by laser light, the detected fluorescence for this particle is indeed visibly lower than in {\bf A}. The difference in brightness is due to the light screening effect introduced when the carbon cap is facing downwards. This is confirmed in {\bf C} where another Janus particle which has sedimented at the bottom surface of the sample chamber is observed to diffuse with the cap-down equilibrium orientation dictated by gravity. Similar to {\bf B}, fluorescence light is screened by the cap, hence the particle appears darker under fluorescence microscopy. Imaging and illumination settings were kept constant for all measurements.
    }
    \label{fig:SI_JanusOrientation}
\end{figure}

\begin{figure}[htb!]
    \centering
    \includegraphics[width=0.6\textwidth]{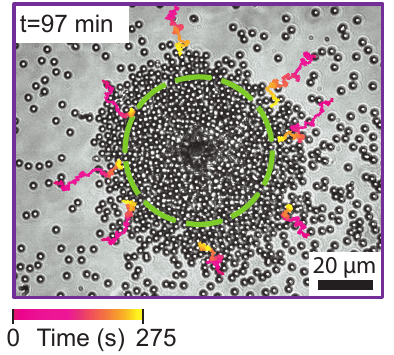}
    \caption{{\bf Reaccumulation of ${\rm \mathbf{TiO_2}}$ colloids after dispersal.} After dispersal (Fig. \ref{fig:Fig1}C), ${\rm TiO_2}$ colloids can be reaccumulated around the Janus particle and lasing action restored (Fig. \ref{fig:Fig1}D). A much shorter time ($\sim$10 minutes) is needed to re-accumulate particles to the same level as at the end of the accumulation phase due to the increased colloidal density after the first round of accumulation and dispersal. For the same reason, lasing action can also be reinstated much faster ($\sim$10 times): lasing is first observed after $\sim$100 s of reaccumulation versus $\sim$1200 s during accumulation. A few 275 second-long trajectories highlight the colloids' motion.
    }
    \label{fig:SI_Reaccumulation}
\end{figure}

\begin{figure}[htb!]
    \centering
    \includegraphics[width=0.6\textwidth]{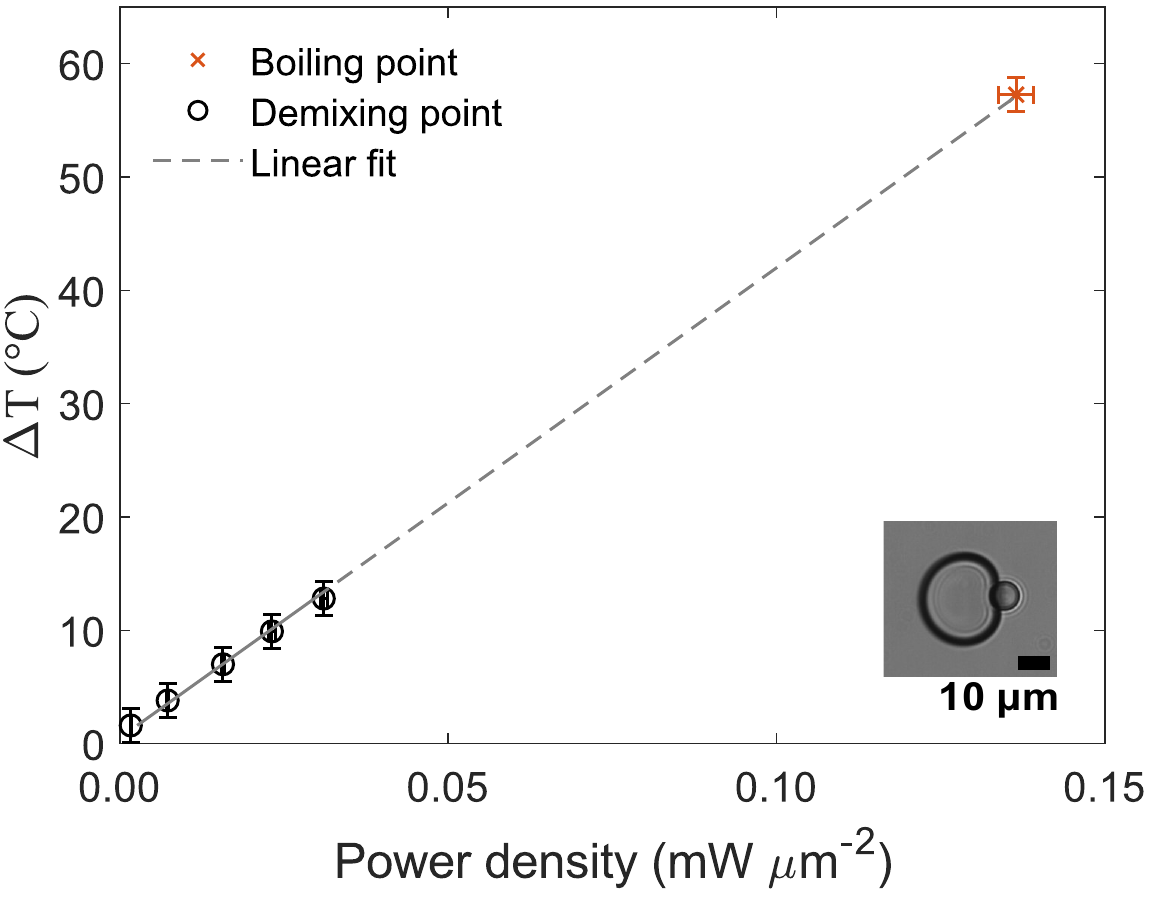}
    \caption{{\bf Temperature increase measurement around a Janus particle.} All our experiments were performed by illuminating a Janus particles at a fixed HeNe laser power density (0.14 mW $\rm \mu m^{-2}$) corresponding to a local temperature increase $\Delta T = 57 \pm 1.6 \, ^{\circ} {\rm C}$ (cross), just below the boiling point of ethanol ($T_{\rm b} = 78.37 \, ^{\circ} {\rm C}$). Above this power density, a cavitation bubble forms around the cap of the Janus particle (inset). To confirm this temperature increase, we calibrated the relationship between laser power density and $\Delta T$ by detecting the onset of the demixing of a critical mixture (2,6-lutidine and water) with respect to a set temperature as a function of laser power (circles). The sample containing the Janus particles for the measurements was placed on a homemade temperature stage with a precision of $0.1 \, {\rm K}$ and allowed to equilibrate for 10 min at a range of different temperatures, before illuminating the particle with gradually increasing laser powers. Each of these temperatures correspond to a well-defined $\Delta T$ with respect to the critical temperature of the mixture ($T_{\rm c} = 307.25 \, {\rm K}$), thus allowing to univocally identify the corresponding power density at which demixing of the critical mixture is clearly observed. Fitting a linear trend to the whole set of data (dashed line) allows us to verify the reliability of our initial temperature estimation due to boiling in ethanol (cross). The bars around each data point represent one standard deviation around the mean values. 
    }
    \label{fig:SI_TMeasurement}
\end{figure}

\begin{figure}[htb!]
    \centering
    \includegraphics[width=0.6\textwidth]{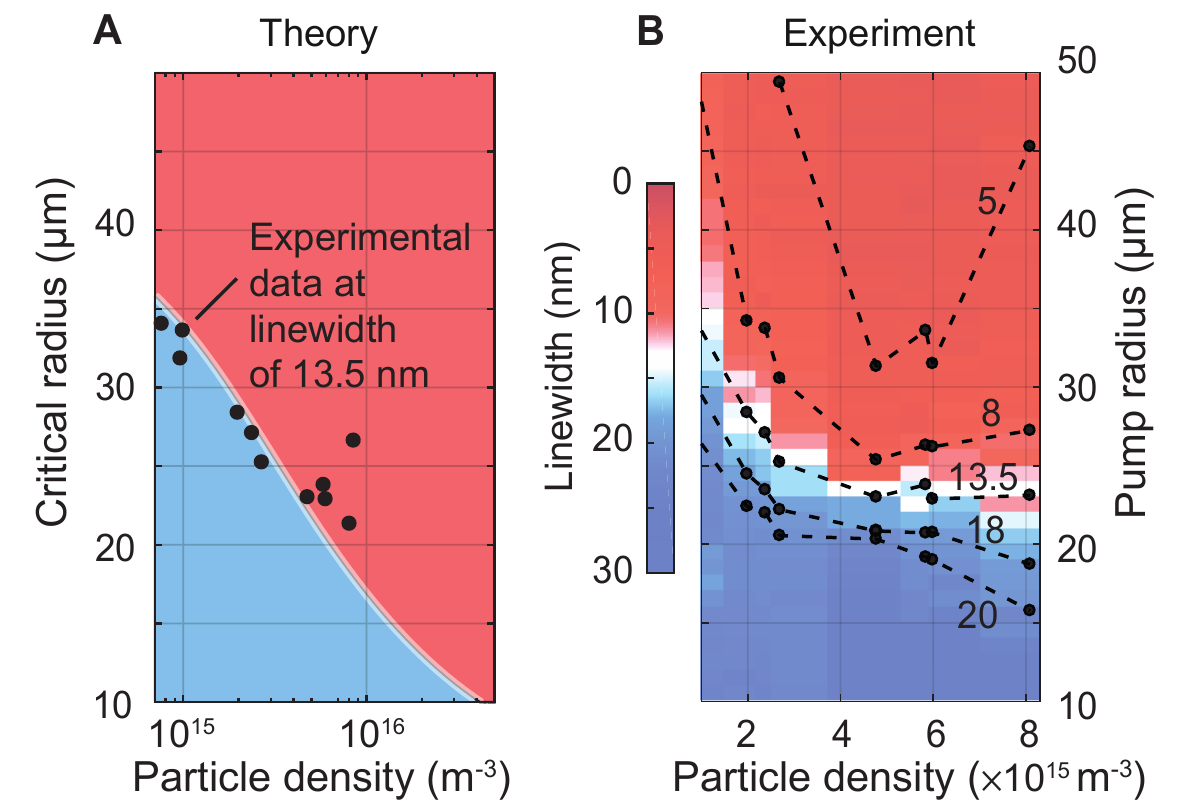}
    \caption{{\bf Characterization of critical radius in ${\rm \mathbf{TiO_2}}$ colloidal samples with rhodamine B dye.} We have performed experiments with both rhodamine 6G and rhodamine B as detailed in the supplementary material. The measurement of the critical radius and its modeling has been performed for both dyes.  Fig. \ref{fig:Fig2}D-E is done with rhodamine 6G, while this figure is done with rhodamine B. In both cases the colloidal samples are similar, and fabricated with the same procedure. The dots in {\bf A} and {\bf B} highlight experimental values  obtained ({\bf A}) at threshold and ({\bf B}) at different linewidths (values in nm in the plot) for different pump radii under a fixed pump fluence (140 mJ cm$^{-2}$). The lasing threshold (white) is defined when the full width at half maximum is 13.5 nm. The RTE model in \textbf{A} uses a gain length of 11.09 $\mu$m as the only free parameter to fit the experimental data \cite{Supplementary}. Theory and experiments show good agreement.
    }
    \label{fig:SI_RhB_Rcr}
\end{figure}

\begin{figure}[htb!]
    \centering
    \includegraphics[width=0.8\textwidth]{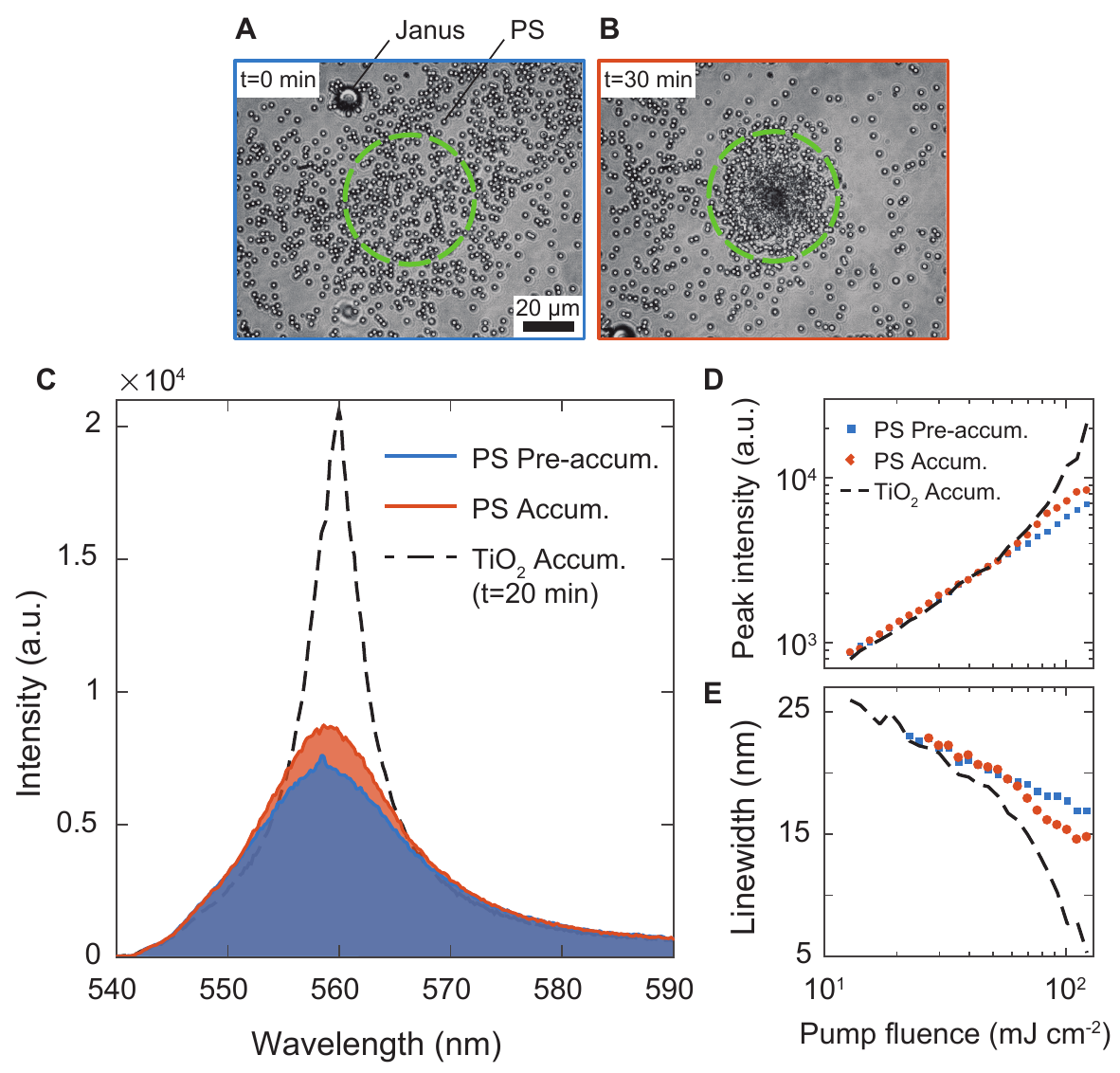}
    \caption{{\bf Absence of lasing in assemblies of low refractive index colloids.} ({\bf A}-{\bf B}) A light-absorbing Janus particle in a laser dye solution with polymer (polystyrene, PS) microparticles attracts the diffusing colloids (accumulation) when illuminated by a  HeNe laser (CW, 632.8 nm). ({\bf B}) The colloids assemble in a dense cluster as in Fig. \ref{fig:Fig1} for TiO$_2$ particles. The polystyrene colloids have similar size and initial concentration as the TiO$_2$ particles in Fig. \ref{fig:Fig1}. The dashed area in {\bf A}-{\bf B} represents the pump region (52 $\mu$m in diameter). ({\bf C}) As scattering from the polystyrene particles is weaker than that from TiO$_2$ particles due to their lower refractive index, the spectra for the polymer colloids before and after accumulation show no lasing action. Nonetheless, a similar size cluster of TiO$_2$ particles is lasing (black dashed line). All spectra are obtained at the pump fluence of 130 mJ~cm$^{-2}$. The absence of lasing action from the polystyrene cluster is confirmed by the linear (rather than superlinear) increase of the peak intensity as a function of pump fluence in {\bf D} and by the fact that the emission linewidth does not reduce below 13.5 nm (lasing threshold) in {\bf E}.
    }
    \label{fig:SI_NoLasing}
\end{figure}

\end{document}